\documentclass[aps,prl,twocolumn,groupedaddress,superscriptaddress]{revtex4-1}
\usepackage{amsmath,amsbsy, graphicx, textcomp, tikz, upgreek, hyperref, subfigure}
\usepackage{color, todonotes}
\begin{document}

\makeatletter
\newcommand*{\balancecolsandclearpage}{%
  \close@column@grid
  \clearpage
  \twocolumngrid
}
\makeatother


\title{Propulsion Mechanisms for Leidenfrost Solids on Ratchets}

\author{Tobias Baier}
\email[]{baier@csi.tu-darmstadt.de}
\affiliation{Center of Smart Interfaces, TU Darmstadt, Germany}

\author{Guillaume Dupeux}
\email[]{guillaume.dupeux@espci.fr}
\affiliation{PMMH, UMR 7636 du CNRS, ESPCI, 75005 Paris, France}
\affiliation{Ladhyx, UMR 7646 du CNRS, Ecole Polytechnique, 91120 Palaiseau, France}

\author{Stefan Herbert}
\email[]{herbert@ttd.tu-darmstadt.de}
\affiliation{Technische Thermodynamik, TU Darmstadt, Germany}

\author{Steffen Hardt}
\email[]{hardt@csi.tu-darmstadt.de}
\affiliation{Center of Smart Interfaces, TU Darmstadt, Germany}

\author{David Qu\'er\'e}
\email[]{david.quere@espci.fr}
\affiliation{PMMH, UMR 7636 du CNRS, ESPCI, 75005 Paris, France}
\affiliation{Ladhyx, UMR 7646 du CNRS, Ecole Polytechnique, 91120 Palaiseau, France}

\date{\today}

\begin{abstract}
We propose a model for the propulsion of Leidenfrost solids on ratchets based on viscous drag due to the flow of evaporating vapor. The model assumes pressure-driven flow described by the Navier-Stokes equations and is mainly studied in lubrication approximation. A scaling expression is derived for the dependence of the propulsive force on geometric parameters of the ratchet surface and properties of the sublimating solid. We show that the model results as well as the scaling law compare favorably with experiments and are able to reproduce the experimentally observed scaling with the size of the solid.
\end{abstract}



\maketitle

The propulsion mechanism of self propelling Leidenfrost drops on ratchets has been debated since the first publication of the phenomenon by Linke et al. in 2006 \cite{Linke_2006}. A simpler system that leaves out many of the complications due to the deformable nature of the drops \cite{Dupeux_2011, Cousins_2012, Ok_2011, Feng_2012}, are Leidenfrost solids. Similar to drops, a platelet of dry ice levitating on a cushion of its own vapor over a hot ratchet surface  shows directed motion \cite{Lagubeau_2011}. Even for such a simplified system a debate is going on about what the physical mechanism responsible for propulsion could be. It has been suggested that the Leidenfrost solid is driven by a "rocket effect" originating from the recoil of the vapor produced from sublimation \cite{Lagubeau_2011}. On the other hand, an attempt to explain the propulsion via thermal creep flow due to rarefaction effects in the vapor phase has been published \cite{Wuerger_2011}. Finally, a viscous mechanism due to pressure-driven flow has been suggested \cite{Linke_2006, Cousins_2012, Dupeux_2011b}. As sublimated vapor is drained sideways along the grooves, perpendicular to the movement of the Leidenfrost solid, it is additionally directed down the slope of the ratchet towards the deep sections of the groove, dragging the levitated solid along \cite{Dupeux_2011b}.

In this paper we corroborate the picture of pressure-driven viscous flow as the main propulsion mechanism for typical scenarios studied in experiments. Our work is based on a combination of model calculations and experimental studies. We propose a model based on a continuum description for the velocity, pressure and temperature fields in the gap between the hot surface and the dry ice. The propulsive force is obtained from the viscous drag on the surface of the Leidenfrost solid, and we develop a scaling expression reflecting how this force depends on geometric parameters of the ratchet surface and properties of the sublimating solid. As such, it supersedes the scaling analysis given in \cite{Dupeux_2011b} by taking into account the 3-D character of the vapor flow, allowing us to understand the series of experiments also presented in the paper.

For our model we assume that the gas velocity $u_i$ and pressure $p$ as well as the temperature distribution $T$ between the hot surface and the platelet of dry ice are governed by the steady state incompressible Navier-Stokes equations
\begin{equation}\label{eq:NS}
  \rho u_j \partial_j u_i = \partial_j \{\eta (\partial_i u_j+\partial_j u_i) - \delta_{ij} p\},
\end{equation}
with $\partial_i u_i =0$ and the energy equation
\begin{equation}\label{eq:HE}
  \rho c_p u_i \partial_i T = \partial_i (\lambda \partial_i T).
\end{equation}
For simplicity we assume the vapor properties to be constant. In particular we will use a density of $\rho=1.2 \,\mathrm{kg/m^3}$, a viscosity $\eta=22 \,\mathrm{\upmu Pa\,s}$, heat capacity $c_p=980 \,\mathrm{J/(kg\, K)}$ and thermal conductivity $\lambda=29 \,\mathrm{mW/(m\!\cdot\! K)}$, corresponding to carbon dioxide at a temperature of $180^\circ \mathrm{C}$ and a pressure of 1 bar.

Further, we assume the no-slip boundary condition at the surfaces of the ratchet and the dry ice. The sublimation rate is determined by equating the heat flux normal to the wall $J_q=-\lambda \partial_n T$ with the heat sink due to sublimation, such that the normal velocity at the ice surface becomes $u_n=\lambda \partial_n T/(\rho\,\Delta H_\mathrm{subl})$, where the subscript $n$ designates the normal component and $\Delta H_\mathrm{subl}=598\,\mathrm{kJ/kg}$ is the latent heat of sublimation. We assume the ice surface to be at saturation temperature, $T_\mathrm{subl}=-78.6^\circ\mathrm{C}$, and the ratchet surface to be at a constant temperature of $T_0=450^\circ\mathrm{C}$.  Despite the inhomogeneous sublimation, we also consider that the ice surface remains flat. Since the platelet moves across the ratchet and hence the surface samples over regions of different sublimation rate, this assumption should be reasonable. 

\begin{figure}
\includegraphics[scale=0.4]{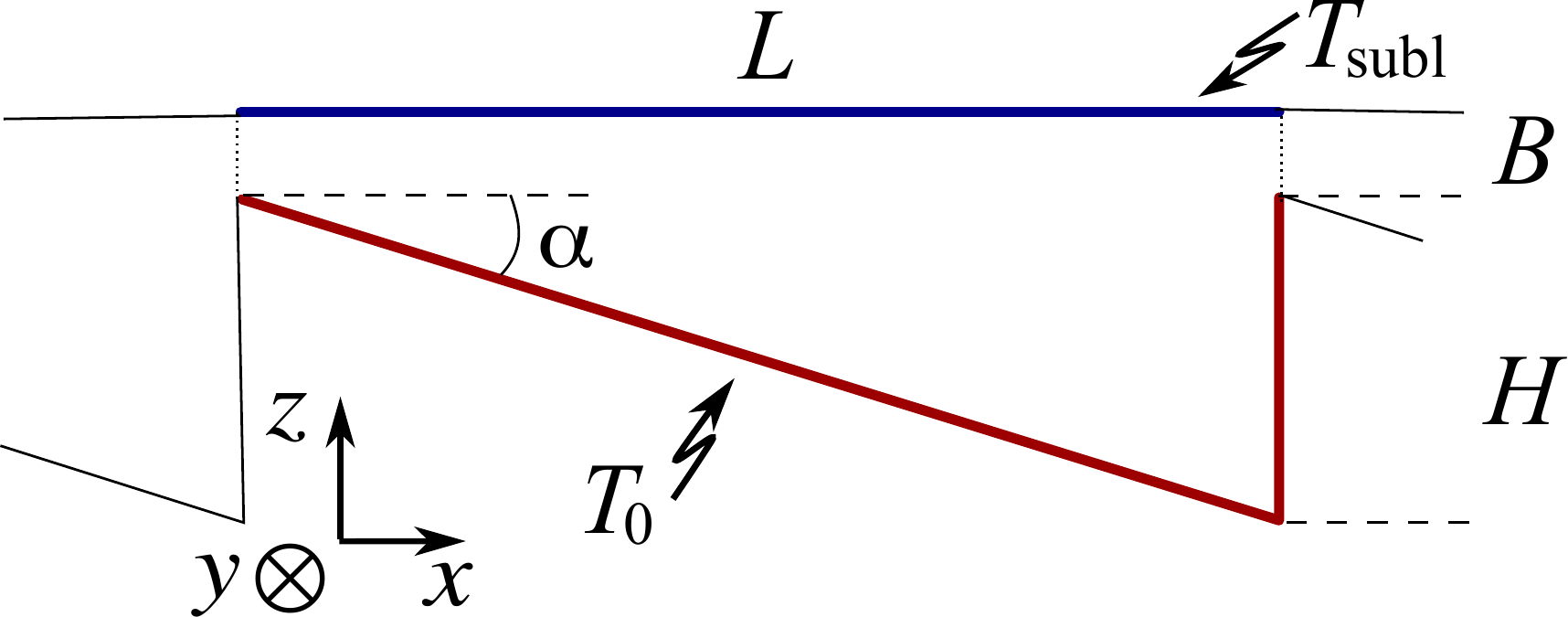}
\caption{\label{fig:Geom} (Color online) Sketch of the geometry considered. The dry ice at temperature $T_\mathrm{subl}$ hovers a distance $B$ above the ratchet geometry. The ratchet grooves have periodicity $L$ and depth $H$. In our modeling we consider a single groove of $y$-extension $2R$ with periodicity in $x$-direction, as well as a disc of radius $R$, covering several grooves.}
\end{figure}

The main parameters of interest are the average pressure $\bar p$ and shear stress $\tau\equiv\bar\tau_{zx}$ at the ice surface. The first determines which weight of the ice block can be supported by the vapor cushion, and the second the viscous propulsion force. Using the density $1560\,\mathrm{kg/m^3}$ of dry ice, the pressure can be converted into a platelet thickness. From the evaporation rate it is apparent that the normal velocity scale is governed by $U_0=(\lambda\Delta T/\Lambda)/(\rho \Delta H_\mathrm{subl})$, where $\Delta T$ is the temperature difference between the surfaces and $\Lambda$ is some length scale of the order of the gap width between ratchet and ice, c.f. figure \ref{fig:Geom}. Note in particular that the corresponding Reynolds number $\mathrm{Re}=\rho U_0\Lambda/\eta$ is independent of this length scale. With the above parameters we get $\mathrm{Re}\approx 1$. Correspondingly, since the forces are assumed to be viscous, the relevant scale for the stresses is $\Pi_0\sim \eta U_0/\Lambda \sim \eta\lambda\Delta T/(\Lambda^2\rho\Delta H_\mathrm{subl})$, containing all of the material properties of the vapor and the temperature difference. 

To start with, consider only a single groove of width $L$ and depth $H$ (c.f. figure \ref{fig:Geom}). This situation can be thought of as a platelet that hovers over many grooves, of which one is studied. An ice block of width $2R$ is assumed to levitate at a distance $B$ from the surface. Further, we assume the flow, pressure and temperature fields to be periodic in the direction normal to the groove ($x$-direction). At the sides of the ice-block ($y=\pm R$) the pressure is ambient and the normal derivatives of temperature and velocities vanish. Since $R\gg L,\,H,\,B$, the details of this boundary condition do not strongly affect the flow inside the groove. 

\begin{figure}
\includegraphics[scale=0.097, clip=true, trim=0cm 0cm 0cm 0cm]{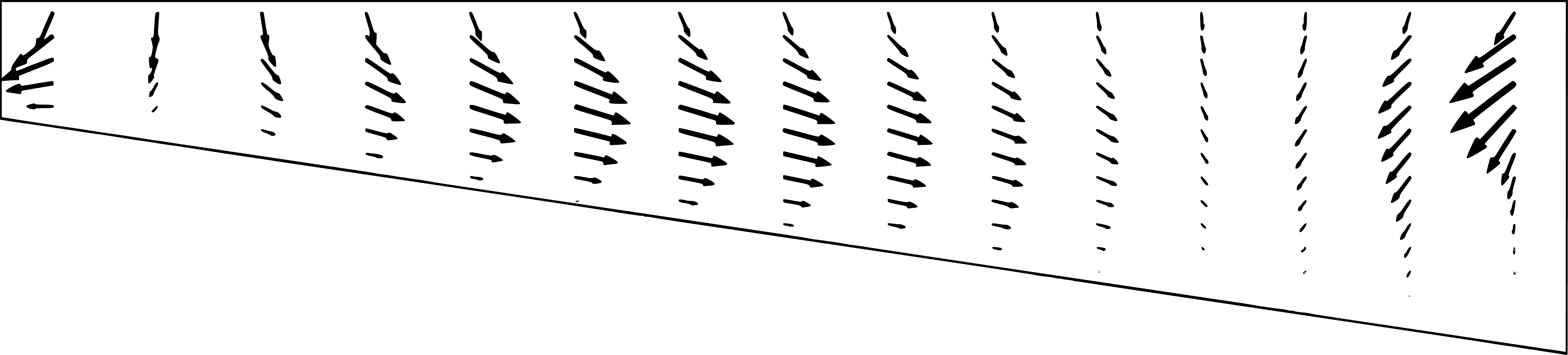}
\includegraphics[scale=0.097, clip=true, trim=0cm 0cm 0cm 0cm]{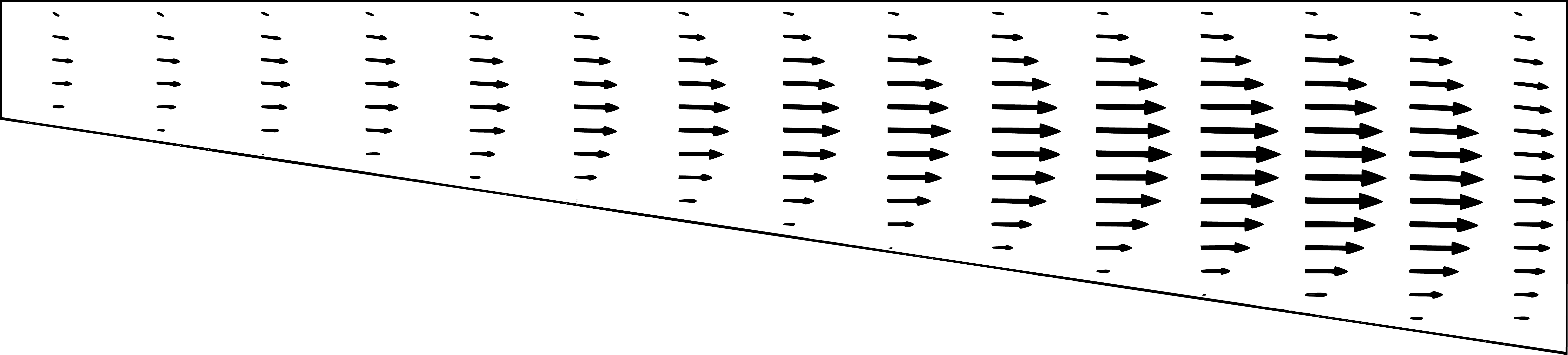}
\caption{\label{fig:Shear} Projected velocity profiles at $y=R/2$, both $xz$- (top) and $yz$-velocities (bottom); since $u_y\gg u_x$ the arrow scales are not identical (differing by a factor $\mathcal{O}(10)$ at this position). $R=5\,\mathrm{mm}$, $L=1\,\mathrm{mm}$, $H/L=0.15$, $B=75\,\mathrm{\upmu m}$. }
\end{figure}

As an example, we consider the parameters $R=5\,\mathrm{mm}$, $L=1\,\mathrm{mm}$, $H/L=0.15$ and $B=75\,\mathrm{\upmu m}$. We solve equations (\ref{eq:NS}) and (\ref{eq:HE}) via finite element discretisation as implemented in Comsol Multiphysics 4.1. The structure of the solution is as follows. The main pressure gradient is along the groove, underpinning that the groove itself acts mainly as a drain for the vapor. The temperature profile and the gradients in pressure normal to the $y$-direction become essentially independent of $y$. Figure \ref{fig:Shear} shows the flow field projected onto the plane normal to the groove (arrows), at position $y=R/2$.  Again, this distribution is almost independent of $y$. The flow field is essentially as described above: the strong flow along the groove is fed by the regions of strong sublimation at the narrow gap. The velocity field is almost parabolic, albeit slightly squeezed in the $xz$-projection due to the evaporation mass flux at the top wall. 

Since a direct computation of the flow below a realistically large platelet is beyond our reach, we propose a simplified model, assuming -- in the spirit of a lubrication approximation -- a parabolic flow profile everywhere, neglecting all inertial terms. Defining the average flow rate in $i$ direction ($i\in \{x,y\}$) by integrating over the height
\begin{equation}
  Q_i(x,y)=\int_0^{h(x,y)} dz\, u_i(x,y,z),
\end{equation}
where $h(x,y)$ is the local distance between the ratchet and ice surfaces, the Reynolds (lubrication) equation becomes
\begin{equation}
  \partial_i p(x,y) = -\frac{12\eta}{h^3(x,y)}Q_i(x,y).
\end{equation}
Mass conservation dictates $\partial_i Q_i = u_n$, with a source term due to sublimation equal to the normal velocity at the ice surface as defined before. In the spirit of the lubrication approximation we assume the local temperature gradient to be proportional to $1/h(x,y)$. Taking the divergence of the Reynolds equation then yields
\begin{equation}\label{eq:ReynoldsDim}
  \partial_i(h^3\partial_i p)=-12 (\Pi_0 \Lambda^2)/h.
\end{equation}

We solve equation (\ref{eq:ReynoldsDim}) in the case of a single groove, assuming periodicity ($p(x,y)=p(x+L,y)$), as well as in the case of a circular platelet. As boundary condition at the edge of the domain $\Omega$, i.e. the circumference of the platelet, we assume $p|_{\partial\Omega}=0$ and neglect any outlet effects beyond that. For the solution a finite element discretisation as implemented in Comsol Multiphysics is used. In our approximation, the wall shear stress is given by $\boldsymbol{\tau}_w=-\frac{1}{2}h\boldsymbol{\nabla}p$.

\begin{figure}
\includegraphics[width=8cm]{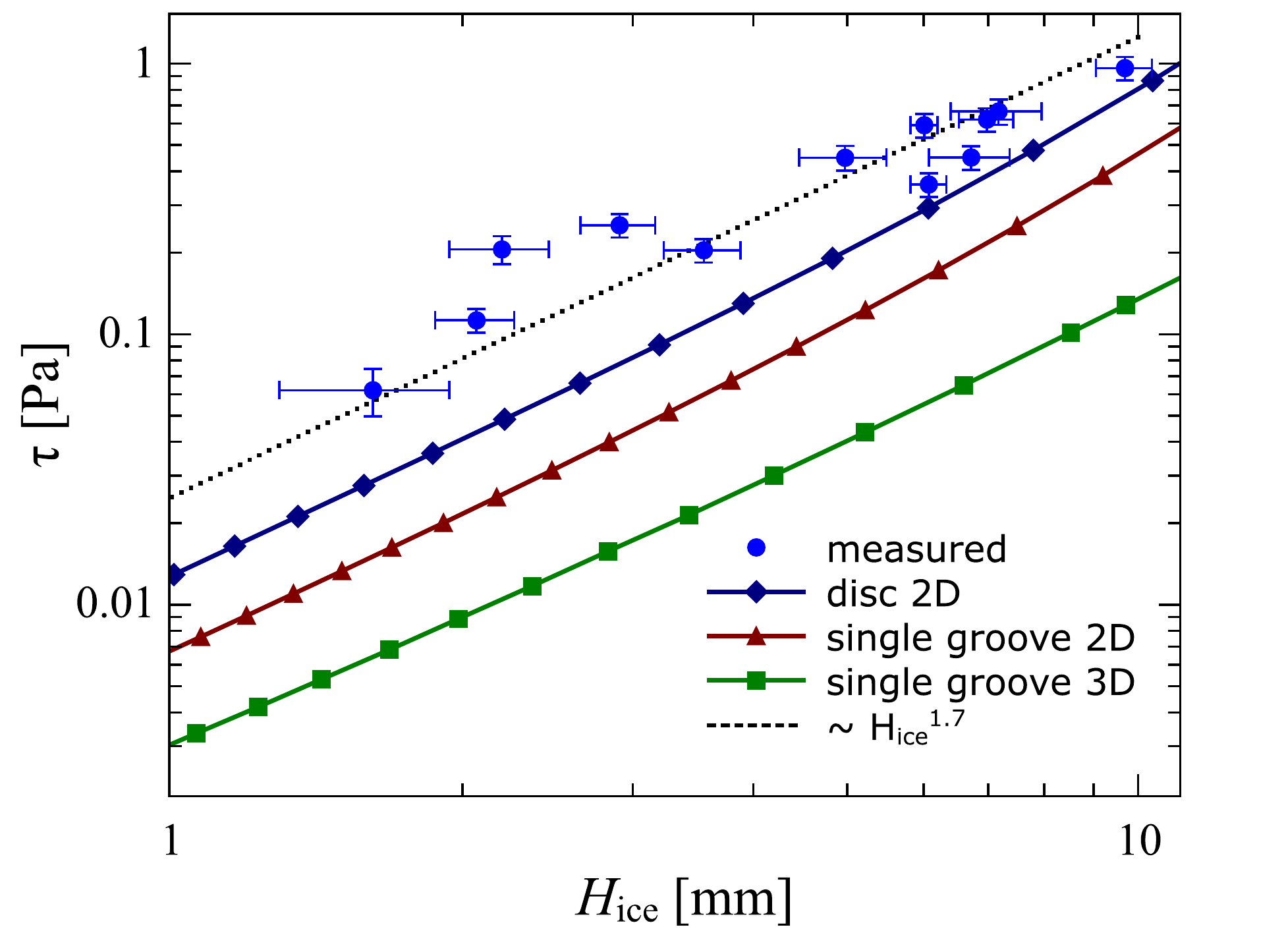}
\caption{\label{fig:tau_vs_HIce} (Color online) Average shear stress $\tau$ vs. thickness of ice. Experimental data (blue circles), 2D lubrication model (disc: dark blue diamonds; single groove: red triangles), 3D simulation (single groove: green squares). The dotted line shows $\tau\sim H_\mathrm{ice}^{1.7}$. $L=1.5\,\mathrm{mm}$, $H=0.25\,\mathrm{mm}$, $R=7\,\mathrm{mm}$.}
\end{figure}

The lubrication model and the 3D model for a single groove give similar shear stress distributions at the upper wall. In both cases a net shear stress is generated driving the platelet in positive $x$-direction. To verify the model predictions, experiments with levitated dry ice platelets have been performed. A quantity that is directly accessible experimentally is the acceleration and hence the net force on a dry ice platelet of known mass. The stress simply is this force divided by $\pi R^2$. For this, cylinders of dry ice with a diameter of $2R=14\,\mathrm{mm}\pm 0.6\mathrm{mm}$ and of different heights are placed on a ratchet heated to $450\,^\circ\mathrm{C}$. Subsequently, the propelling force is measured using the method described in \cite{Dupeux_2011b} (c.f. supplemental material). The geometry parameters of the ratchet are $L=1.5\,\mathrm{mm}$ and $H = 0.25\,\mathrm{mm}$. In figure \ref{fig:tau_vs_HIce} the experimentally obtained average wall shear stress is plotted vs. the platelet thickness for values between $1\,\mathrm{mm}$ and $10\,\mathrm{mm}$. In the same diagram the corresponding values obtained with the 3D model for a single groove and with the lubrication model both for a single groove and a circular disc are displayed. The model results are unanimously compatible with a scaling $\bar\tau_{zx}\sim H_\mathrm{ice}^{1.7}$ of the shear stress with the thickness of the platelet, agreeing well with the experimental data. However, the different models predict slightly different values for the shear stress at a given thickness. For the single groove, the lubrication model predicts a slightly larger shear stress compared to the more complete 3D model. This is partly due to the neglect of inertial effects and partly due to the simplified temperature profile used in the lubrication model. Comparing the lubrication results obtained for a single groove and the more adequate circular disc, the latter predicts a larger shear stress. This is due to the fact that for the disc a portion of the vapor can also escape to the front and back. Therefore the disc hovers closer to the surface, resulting in a larger sublimation rate and shear stress $\tau$. This same effect would also play a role in a more complete 3D model of a full disc, raising the model results closer to the experimental data. We conclude that a model based on viscous forces due to the sublimation mass flux not only gives the correct order of magnitude for the forces on a Leidenfrost solid but is able to reproduce the scaling of the forces with respect to the height of the platelet.

\begin{figure}
\includegraphics[width=8cm]{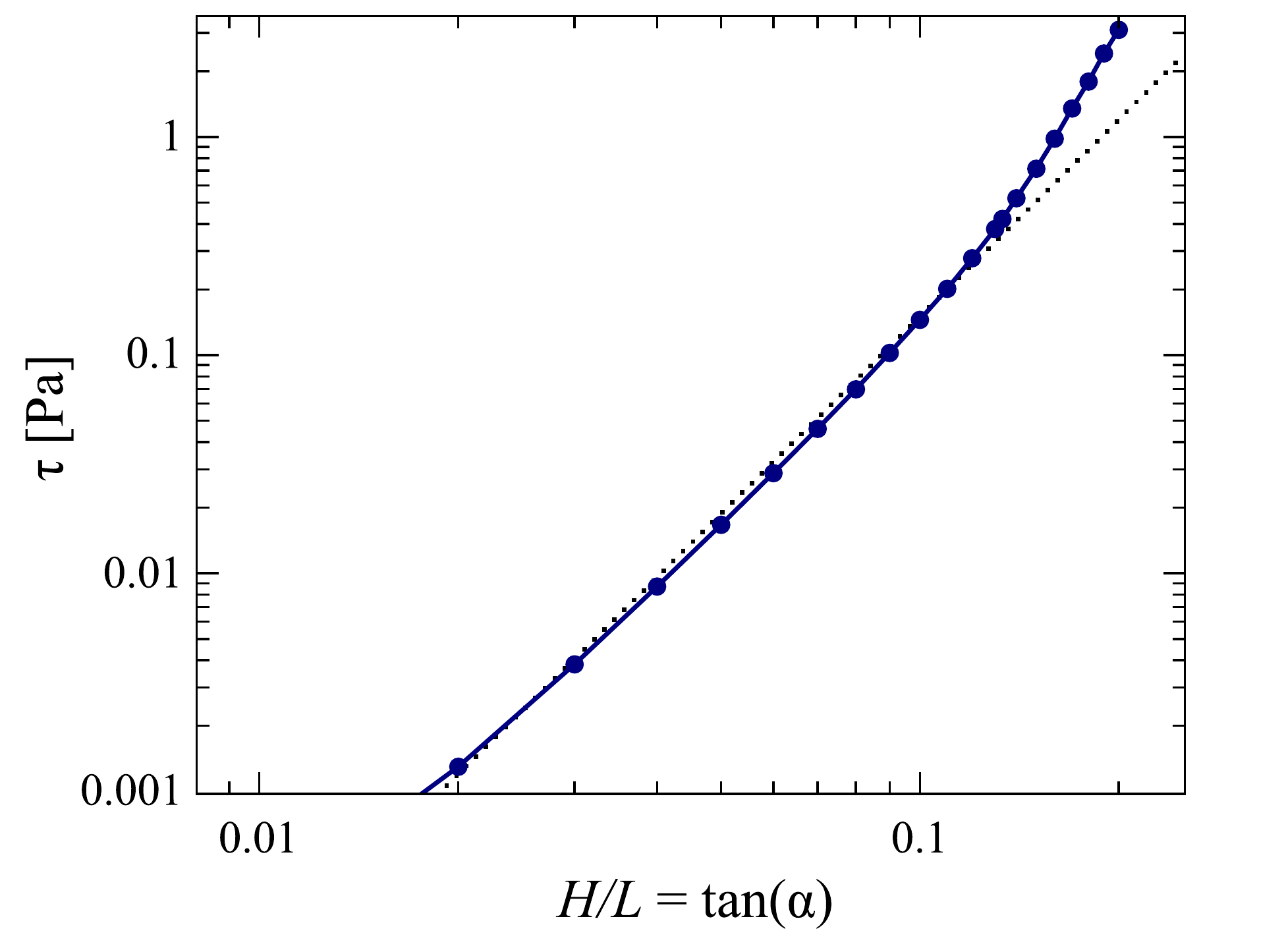}
\caption{\label{fig:tau_vs_tanA} (Color online) Average shear stress $\tau$ for a disc in lubrication approximation as function of the groove angle. $\tan\alpha=H/L$ is varied at fixed $L=1.5\,\mathrm{mm}$ and $R= 5\,\mathrm{mm}$. $B$ is chosen such that the average pressure below the disc becomes $\bar{p} = 100\,\mathrm{Pa}$. The dotted line shows $\tau\sim \tan^3\alpha$.}
\end{figure}

Based on that, we extend the discussion to the scaling with respect to the geometric parameters of the ratchet. In doing so we assume a constant thickness of the ice platelet such that it is supported by an average gas pressure (above ambient pressure) of $\bar p=100\,\mathrm{Pa}$, which corresponds to a platelet of thickness of $H_\mathrm{ice}\simeq 6.4\,\mathrm{mm}$. The Reynolds equation is solved for a circular platelet, and $B$ is varied till the average pressure is $\bar p$. Figure \ref{fig:tau_vs_tanA} gives an example for the obtained scaling with $\tan\alpha=H/L$. Overall, the scaling with geometric parameters is roughly $\tau(L,H,R)\sim L^1 H^3 R^{-3}$. A similar scaling is obtained when considering a single groove in lubrication approximation or in the 3D model, albeit with slightly different exponents, varying by about 10\%. Note that for large angles the curve for $\tau$ vs. $\tan\alpha=H/L$ deviates from the scaling function, and the force increases more rapidly. In this region $B$ becomes very small, and eventually the pressure below the disc is unable to support its weight, i.e. touchdown occurs. In our experiments no movement of the discs was observed when increasing the angle to $\tan\alpha\simeq 0.29$, indicating touchdown. Note that a similar deviation from the scaling law would be observed for small $R$ for the same reason as for large $\tan\alpha$.

\begin{figure}
\centering
\includegraphics[width=8cm]{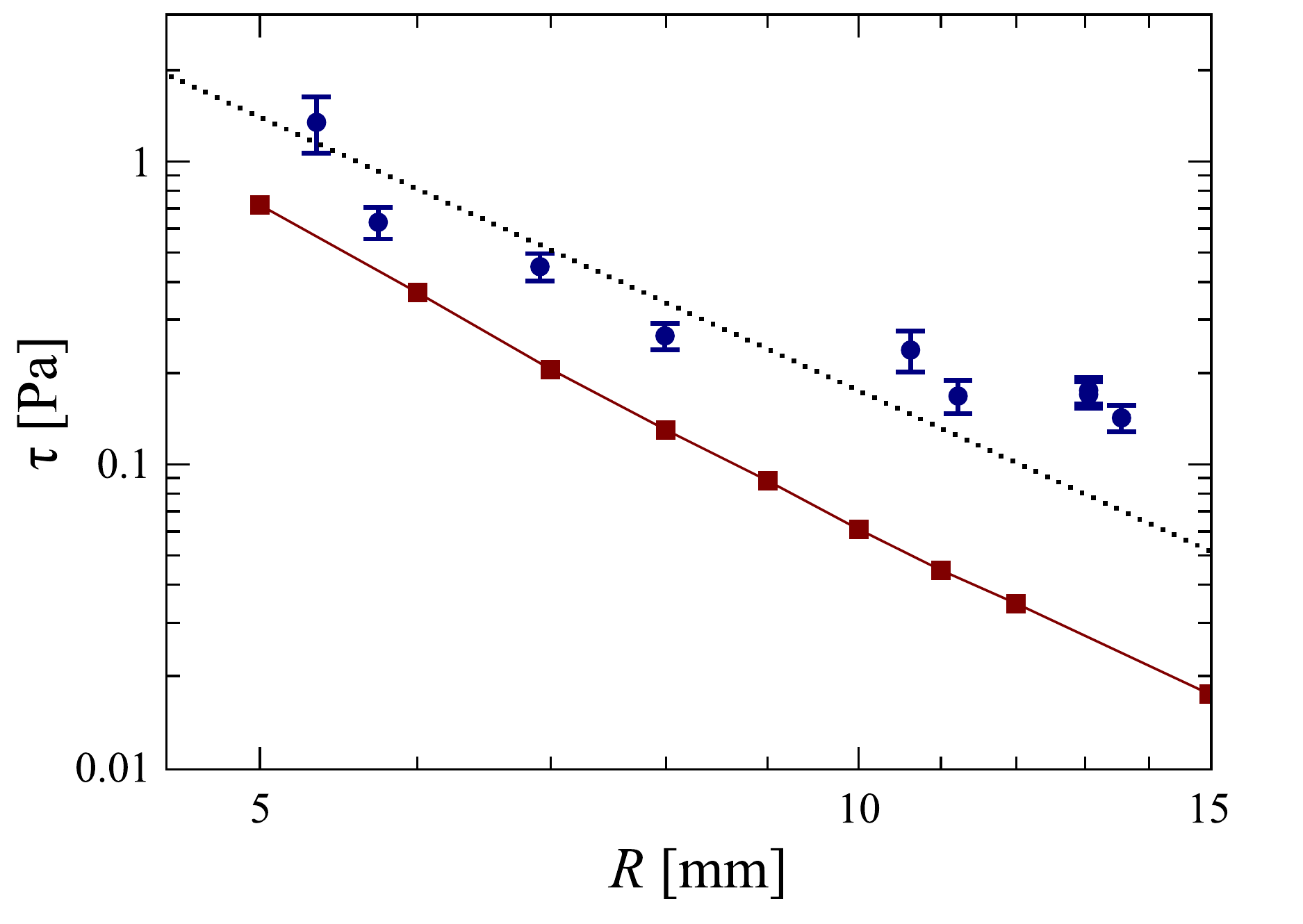}
\caption{\label{fig:Tau_R} (Color online) Average shear stress $\tau$ as a function of the platelet radius $R$. A platelet of height $H_{\mbox{\footnotesize ice}} =$~5~mm and radius $R=7\,\mathrm{mm}$ is considered. Blue circles are experimental results, red squares are results obtained in the 2D lubrication model for a disc of corresponding dimensions. The dotted line shows $\tau\sim R^{-3}$. $L=1.5\,\mathrm{mm}$, $H=0.25\,\mathrm{mm}$.}
\end{figure}

To corroborate the findings of the scaling analysis, we show in figure \ref{fig:Tau_R} the mean stress $\tau$ measured on a platelet as a function of its radius, for a height $H_{\mbox{\footnotesize ice}} =$~5~mm. Corresponding results obtained in the lubrication approximation for a disc are shown in the same figure. The stress decreases as the radius increases, and the dashed line of slope -3 shows a fair agreement with experiments and model results.

It is interesting to understand this behavior (and more generally the scalings found in this study) based on a simple intuitive picture. We divide the flow in a single groove into two parts. Firstly, the vapor flows down the incline (along the $x$-axis in fig.1). In the lubrication approximation, the velocity of this flow is $U_1 \sim (\Lambda^2/\eta) (\Delta P_1/L)$ where $\Lambda$ is a typical vertical length scale, and $\Delta P_1$ is deduced from mass flux conservation (c.f. equation (\ref{eq:ReynoldsDim})): $\Delta P_1 \sim \eta \lambda \Delta T L^2/(\Lambda^4 \rho \Delta H_{\mbox{\footnotesize subl}})$. The $\Delta P_1 \sim L^2$ scaling stems from the quadratic pressure increase with length in a channel with constant inflow across permeable walls. Secondly, the flow escapes laterally (along the $y$-axis in fig.\ref{fig:Geom}) with a velocity $U_2 \sim (H^2/\eta) (\Delta P_2/R)$. The pressure associated with this flow balances the weight of the platelet ($\Delta P_2 \sim \rho_{\mbox{\footnotesize ice}} g H_{\mbox{\footnotesize ice}}$ where $\rho_{\mbox{\footnotesize ice}}$ is the dry ice density). 

The shear stress arises from the "first" flow, so that it scales as $\eta U_1/\Lambda$. We deduce the length $\Lambda$ from mass conservation between the two flows: $U_1 \Lambda R \sim U_2 \Lambda L$. Eventually, we get an expression for $\tau$ which is:
\begin{equation}\label{eq:scaling}
	\tau \sim \left( \frac{\rho \Delta H_{\mbox{\footnotesize subl}} g^3 \rho_{\mbox{\footnotesize ice}}^3}{\eta \Delta T \lambda} \right)^{1/2} \frac{H_{\mbox{\footnotesize ice}}^{3/2}LH^3}{R^3}.
\end{equation}
We find that the stress should vary as $H_{\mbox{\footnotesize ice}}^{1.5}$, $R^{-3}$, $H^3$ and $L$, scalings very close to the ones derived from the lubrication approximation (only $H_{\mbox{\footnotesize ice}}^{1.7}$ deviates slightly). Note that this scaling of equation (\ref{eq:scaling}) is also obtained in the lubrication approximation for a single groove in the limit of small angles (c.f.  supplemental material).

The experimental results and the model both reveal a decrease of the propelling force with the radius of the platelet ($F\sim \tau R^2 \sim R^{-1}$). This behavior stands in contrast to what can be observed for Leidenfrost droplets self-propelling on ratchets, for which the force was shown to increase rapidly with the radius \cite{Lagubeau_2011}. The systems are indeed quite different: a drop on a hot ratchet follows the asperities of the texture, so that the characteristic length scales involved in the scaling laws above get modified --- changing the geometry in these systems has a deep implication on the resulting force.

Based on the observations above, we conclude that viscous drag from pressure driven flow due to sublimation seems the most important driving force for self-propelling Leidenfrost solids on ratchets. However, despite the excellent agreement with scaling predictions our model results consistently lie below the measured values, indicating that some detail may not be fully captured yet. Also note that these findings do not necessarily carry directly over to Leidenfrost drops for which additional effects such as deformability or interfacial stresses come into play.

Contrasting this viscous mechanism based on pressure-driven flow, it was recently suggested that rarefaction effects, in particular thermal creep flow, could play a dominating role for the propulsion on Leidenfrost ratchets \cite{Wuerger_2011}. We have investigated these effects using Monte Carlo simulations in \cite{Hardt_2012}. For typical geometries used in experiments it is shown that such rarefaction effects based on the finite mean free path of gas molecules only play a minor role.

\begin{acknowledgments}
We kindly acknowledge support by the German Research Foundation (DFG) through the Cluster of Excellence 259.
\end{acknowledgments}

\bibliography{litRatchet}

\balancecolsandclearpage
\appendix
\section*{Supplementary material}
\section{Lubrication approximation: single groove}

For a single groove an approximate analytical solution of the Reynolds (lubrication) pressure equation (\ref{eq:ReynoldsDim}),
\begin{equation*}
\partial_i(h^3\partial_i p)=-12 (\Pi_0 \Lambda^2)/h,
\end{equation*}
can be obtained by superposing two pressure fields $p(x,y)=p_x(x)+p_y(y)$. This approximation works well since, apart from close to the outlet, the $x$- and $y$- components of the pressure gradient are almost independent of each other. The pressure equation then becomes separable and we can obtain a closed form solution. In particular, we obtain for the average pressure and the average shear stress (using $\Lambda=L$ as our length-scale and $\Pi_0 = \eta\lambda\Delta T/(L^2\rho\Delta H_\mathrm{subl})$ as pressure scale):
\begin{equation}\label{eq:lubPressSingle}
  \frac{\bar p}{12\Pi_0 L^2} = \frac{4R^2}{3} \frac{\log H_1/B}{(H_1^4-B^4)},
\end{equation}
\begin{align}\label{eq:lubShearSingle}
  &\frac{\bar \tau}{12 \Pi_0 L^3} = \frac{1}{4 H B H_1} \\\nonumber
  &\;\;\;\; -\left( \frac{1}{2H^2(H_1+B)} + \frac{H}{6(H_1^4-B^4)} \right) \log H_1/B,
\end{align}
where $H_1=H+B$. Note that since $\Pi_0\sim L^{-2}$, $\bar p$ becomes independent of $L$, while $\bar\tau$ is linear in $L$ in this approximation.

For $H\ll B$ equations (\ref{eq:lubPressSingle}) and (\ref{eq:lubShearSingle}) can be expanded to lowest order in $H$. Solving the expansion of (\ref{eq:lubPressSingle}) for $B$ and inserting it into the expansion of (\ref{eq:lubShearSingle}) yields the lowest order contribution to the average shear stress as
\begin{align}
\bar \tau &= (12 \Pi_0 L^3) \frac{7\sqrt{3}}{120}\left(\frac{H \sqrt{\bar p/(12\Pi_0 L^2)}}{R} \right)^3 \nonumber\\
&\sim \left(\sqrt{\frac{\rho_{ice}^3g^3}{\Pi_0 L^2}}\right) H_{ice}^{3/2} L H^3 R^{-3},
\end{align}
where we have used $\bar p=\rho_{ice}gH_{ice}$. This exactly corresponds to the scaling found in equation (\ref{eq:scaling}) in the main text.

\section{Experimental procedure}

The experiments were performed on the basis of the procedure described in \cite{Dupeux_2011b}. In order to measure the propelling force of the ratchet, we push a levitating platelet of dry ice of radius $R$ and mass $M$ with some initial velocity in the direction of negative $x$ (fig. \ref{fig:Geom}). The platelet slows down, stops and accelerates in the direction of the propulsion. We record its motion with a fast camera (200 frames per second). We derive the acceleration from the trajectory $x(t)$. A constant acceleration $x''$ is measured around the point of arrest, from which we deduce the propelling force $F = Mx''$. The stress $\tau$ is defined as $F/\pi R^2$.

The ratchet is made of brass. Its teeth are machined with a milling machine whose head was tilted by 10$^\circ$. The length between two teeth is $L =$~1.5 mm. We use to two different techniques to create the platelet. Either we sublimate a small pellet of dry ice of radius $R =$~7~mm to the required height, or we force a block of dry ice into holes of different radii drilled into a hot plate.


\end{document}